\documentclass[prl,twocolumn,superscriptaddress]{revtex4-1}

\usepackage{amsmath, amssymb, mathrsfs}     
\usepackage{graphicx, graphics}             
\DeclareGraphicsExtensions{.png,.pdf}
\usepackage{subfigure}
\usepackage{url}

\usepackage{color}
\usepackage{soul}

\begin{document}

\title{Classification of Strongly Disordered Topological Wires Using Machine Learning}
\author{Ye Zhuang}
\affiliation{Department of Physics and Institute of Condensed Matter Theory, University of Illinois, Urbana, Illinois 61801, USA}
\author{Luiz H. Santos}
\affiliation{Department of Physics, Emory University, Atlanta, Georgia 30322, USA}
\author{Taylor L. Hughes}
\affiliation{Department of Physics and Institute of Condensed Matter Theory, University of Illinois, Urbana, Illinois 61801, USA}
\date{\today}

\begin{abstract}
In this article we apply the random forest machine learning model to classify 1D topological phases when strong disorder is present. We show that using the entanglement spectrum as training features the model gives high classification accuracy. The trained model can be extended to other regions in phase space, and even to other symmetry classes on which it was not trained and still provides accurate results. After performing a detailed analysis of the trained model we find that its dominant classification criteria captures degeneracy in the entanglement spectrum.
\end{abstract}

\maketitle



\section{Introduction}

Since the discovery of topological insulator phases, the problem of their classification has been an important subject.
A classification table, predicated on the stability of a strong topological phase in the presence of disorder,  was proposed for free fermion systems in the ten Altland-Zirnbauer symmetry classes~\cite{schnyder2008classification,ryu2010topological},
though the table does not determine the explicit phases and phase diagrams of model systems.
Many successful determinations of the phase diagrams of low-dimensional disordered topological phases have been made in symmetry classes A, AIII, and BDI using techniques based on, e.g., entanglement properties, level-spacing statistical analysis, and real-space topological indices ~\cite{prodan2010entanglement,song2014effect,ian2014signatures,ian2014topological,song2014aiii}. In this article we propose the use of a new technique based on the random forest (RF) machine learning model to determine the phase diagrams of models of disordered topological phases. 

Recent work has shown that machine learning can provide a new framework for solving problems in physics. 
In our context, promising developments have been achieved in applying machine learning techniques to classifying phases of matter in condensed matter systems. 
Supervised learning has been used directly in characterizing phases in both classical spin systems ~\cite{torlai2016learning,wang2016discover,carrasquilla2017machine} 
and quantum many-body systems ~\cite{broecker2017machine,ch2017machine,huang2017accelerated,zhang2017quantum,van2017learning}.
Neural networks are the most widely used model to identify phases, especially topological phases of matter, and they are  powerful models that have universal approximation capabilities~\cite{hornik1989feedforward,hornik1991multilayer}.
For example, Chern insulators and fractional Chern insulators can be classified by feeding quantum loop topography into a neural network~\cite{zhang2017quantum}.
Unfortunately, the black-box nature of neural networks makes it hard to interpret the trained models, and it is not easy to extract insightful physical intuition about the system under study. Additionally, the large number of hyper-parameters in a neural network can make it difficult to train.

In this paper, we use the RF method as our machine learning model to detect topological phases with strong disorder.
RF is an ensemble method that is capable of representing complicated functions with much fewer parameters as compared with neural networks, and having more easily interpretable classification criteria once the model is trained ~\cite{breiman2001random,friedman2001elements}.
RF is a collection of decision trees, which can be understood as piecewise constant functions in feature space. 
An individual decision tree cannot make good predictions because, in general, predictions of decision trees have large variance. 
Averaging over decision trees reduces variance, making RF a popular method~\cite{liaw2002classification}.
Some major advantages of RF are that it has few hyper-parameters, and is immune to problems such as over-fitting, collinearity, etc.

To train our RF model we will use the entanglement spectrum (ES)\cite{li2008entanglement} of our physical system as our input data.
We will focus on 1D systems where the ES has been widely used to characterize topological phases, and a robust degeneracy in the entanglement spectrum can serve as a general indicator for a topological phase~\cite{pollmann2010entanglement}.
To benchmark our machine learning model we will consider 1D free-fermion wires having chiral symmetry in the AIII class. We choose this system because the disordered phase diagram of this model has been carefully studied\cite{ian2014topological}. Here we find that
the RF model, trained by the ES data generated from a small fraction of the  phase diagram, can be generalized to the full phase space with high prediction accuracy.
Furthermore, the RF model trained from the AIII class data shows high prediction ability for wires in symmetry class BDI as well.
A detailed analysis reveals that the RF model is primarily capturing the degeneracy in the ES to make its classification, and may provide new routes to identify disordered topological phases from their entanglement properties.

\section{Model}

We start from the disordered chiral Hamiltonian in Ref.~\cite{ian2014topological} defined on a one-dimensional chain with two sites $A$ and $B$ in one unit cell:
\begin{equation}
H = \sum_n \left[\frac{t_n}{2} c^{\dagger}_n \left(\sigma_x+i\sigma_y\right) c_{n+1} + h.c \right], 
+ \sum_n m_n c^{\dagger}_n \sigma_y c_n.\label{eq:chiral}
\end{equation}
where $c^{\dagger}_n=(c^{\dagger}_{n,A}, c^{\dagger}_{n,B})$ are fermion creation operators in unit cell $n$. 
We have included disorder in both the hopping and mass terms, i.e., $t_n=1+W_1\omega_1$, and $m_n=m+W_2\omega_2$, 
where $\omega_1$ and $\omega_2$ are random variables generated from a uniform distribution on $[-0.5,0.5],$ and $W_1, W_2$ represent the strengths of the disorder. The model preserves chiral symmetry $\mathcal{C}H\mathcal{C}^{-1}=-H$ with $\mathcal{C}=\sum_n c^{\dagger}_n \sigma_z c_n$.

In the clean limit, the system has translational symmetry and the Bloch Hamiltonian is
\begin{equation}
\mathcal{H}(k) = t\cos k \sigma_x + (t\sin k + m) \sigma_y.
\end{equation}
The chiral symmetry operator $\chi=\sigma_z$ anti-commutes with $\mathcal{H}(k),$ and one can use the 
topological winding number $\nu$ to identify the $\mathbb{Z}$ classification~\cite{schnyder2009lattice}. 
If we write the Bloch Hamiltonian in the form $\mathcal{H}(k) = d_x(k) \sigma_x + d_y(k) \sigma_y,$ then
$\nu$ is the number of times the unit vector $(\hat{d}_x,\hat{d}_y)$ travels around the origin as $k$ traverses the whole Brillouin Zone. 
For example, when $|m|<|t|$, the system is in symmetry protected topological (SPT) phase with winding number $\nu=1$, and when $|m|>|t|$ the system has $\nu=0.$

When disorder is turned on, and in the limit that $W_2\gg t$, the system is completely dimerized within individual unit cells, i.e., it is in the topologically trivial atomic limit. This result is independent of the value of $m,$ and hence there must be a phase transition when $W_2$ is gradually increased from zero when $|m|<|t|$. 
One signature of the topological phase transition point is the divergence of the localization length of states at the Fermi-level. Using this criterion one can determine an analytic relation that is satisfied at a topological critical point~\cite{ian2014topological}.
\begin{equation}
\frac{|2+W_1|^{1/W_1+1/2}|2m-W_1|^{m/W_1-1/2}}{|2-W_2|^{1/W_2-1/2}|2m+W_2|^{m/W_2+1/2}}=1.\label{eq:transition}
\end{equation} Indeed, using this relation one can determine the phase diagram of this model even in the presence of disorder.

\section{Results}

\begin{figure}
	\centering
	\subfigure[]{\includegraphics[width=0.59\linewidth]{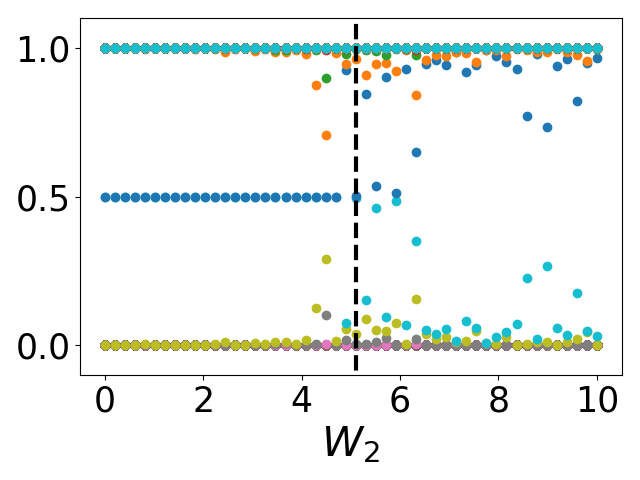}\label{fig:ces}}
	\subfigure[]{\includegraphics[width=0.38\linewidth]{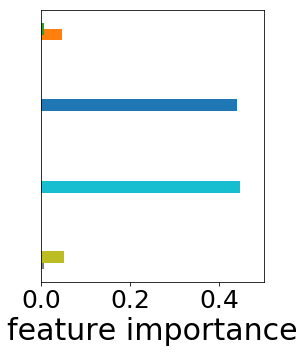}\label{fig:importance}}
	\caption{(a) Single particle entanglement spectrum plotted vs disorder strength $W_2$. 
		The vertical black dashed line is the analytical transition point.
		Double degeneracy at 0.5 on the left hand side indicates the non-trivial SPT phase. 
		There may be accidental degeneracies in the trivial phase due to disorder.(b) Feature importance in our trained random forest model. The vertical axis is the average of the entanglement spectrum for each band in panel (a). We plot the feature importance of the bands in the horizontal direction. The high importance values for the middle two bands, i.e., the two that include the degenerate modes at 0.5, indicate their high influence on model predictions. The bands are colored-coded the same way in panels (a) and (b).}
\end{figure}

To generate training and testing data we calculate the single-particle entanglement spectrum~\cite{ingo2003calculation} using a central spatial cut of the lattice model.
We use periodic boundary conditions on a chain of length $L=400$ with $t=1.$ 
To be explicit, let us first focus on a line in the 3D phase space $\{m,W_1,W_2\}$ with $m=0.5$ and $W_1=1$.
For illustration we plot the ES of one disorder configuration for each value of $W_2$ in Fig.~\ref{fig:ces}. 
The black vertical line indicates the theoretical transition point calculated from Eq.~\ref{eq:transition}. 
We can clearly see double degeneracy at 0.5 in the region of weak disorder, which is a signature for SPT phases~\cite{pollmann2010entanglement}.
In the region of strong disorder, on the other hand, there are no such degeneracies in general,
even though there may be accidental degeneracies induced by disorder.

In order to test the predictive power of the RF model, we first train the model in regions  deep in
the two phases using a set of test data based on the entanglement spectrum. We will further evaluate the RF model by testing if it
can provide accurate predictions for other values of parameters different than those used to 
generate the training data. 
In particular, we test whether the model is capable of detecting the behavior of the system near the 
topological phase transition despite training it deep in the phases. We will indeed verify 
that the RF model can accurately detect disorder-induced phase transitions.

In order to implement the data training,
5000 \emph{training} samples were generated with $W_2$ ranging from 0 to 4 and from 7 to 10, which correspond,
respectively, to the topologically non-trivial and trivial phases.
We intentionally skipped the region near the phase transition point $W_2 \approx 5$, in hopes that the RF model can locate it only with knowledge deep in the phases.
\emph{Test} data was generated separately over the whole range of $W_2$ from 0 to 10, and includes the transition region.
In order to test the performance of the RF model with respect to other widely 
utilized algorithms, we trained three models using the same training and testing data: the linear model (LM), neural network (NN), and random forest,
and we compare the predictions of the first two in relation to the latter. In our numerical analysis, the python package sklearn~\cite{scikit-learn} was used for training and predicting.

\begin{figure*}[t]
	\centering
	\subfigure[]{\includegraphics[width=0.32\linewidth]{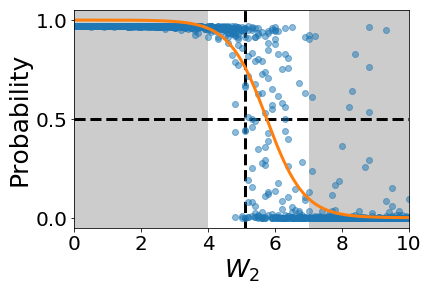}\label{fig:ces_fit_lm}}
	\subfigure[]{\includegraphics[width=0.32\linewidth]{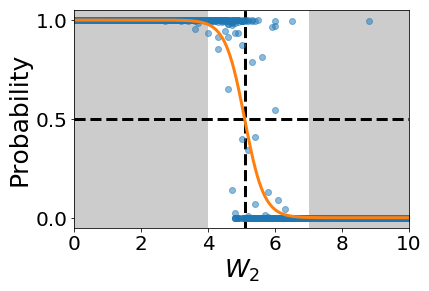}\label{fig:ces_fit_nn}}
	\subfigure[]{\includegraphics[width=0.32\linewidth]{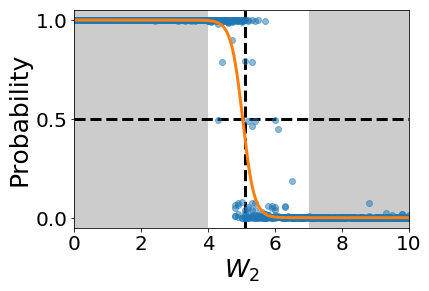}\label{fig:ces_fit_rf}}
	\caption{(a) Predicted probability of being in the topological phase for the three models:
		(a) linear model, (b) neural network, and (c) random forest. The smooth curves are fits using Eq.~\ref{eq:fit}. }\label{fig:ces_fit}
\end{figure*}

We show the prediction results of the three models in Fig.~\ref{fig:ces_fit}. 
 From left to right, the dots in the subfigures represent predicted probabilities of being in the topological phase from LM, NN, and RF models respectively.
We observe that, while most ground states are correctly classified for all the three models, two features stand out. First,
the LM has a number of misclassified states in the region of strong disorder due to the simple linear assumption. 
The trained LM gives a linear relationship between the logarithm of the probability and the gap between the middle two entanglement bands. When the gap is small the model predicts the state is topologically non-trivial with high probability even when it is a trivial state.
Second, while the NN and RF models give reliable predictions for strong and weak values of disorder,
they classify with much higher variance near the phase boundary, which makes it hard to predict the resulting phase from a single disorder configuration. This behavior may be expected for, near the transition,
the (entanglement) spectral gap approaches zero, and the disorder causes strong fluctuations that make it difficult to correctly distinguish the intrinsic degeneracies of the ES on the topological side versus
the frequently encountered accidental degeneracies in the trivial phase near the phase boundary.

The predicted critical point can be obtained for each model by
fitting the predicted probability with the function
\begin{equation} \label{eq:fit}
f(x) = \frac{1}{1+e^{b+wx}}.
\end{equation}
The fitted curves are shown in each of the subfigures in Fig.~\ref{fig:ces_fit}. 
To help identify the phase boundary we choose a cutoff value of 0.5, i.e., when the predicted probability is larger than 0.5, we say the state is in the topological phase; otherwise it is in the trivial phase.
Therefore, the transition happens at the crossing point of the fitted curve and the horizontal dashed line at probability 0.5.
For comparison, the black vertical dashed lines in each subfigure represent the true transition point.
We see that both NN and RF  models can predict the phase transition point with high accuracy, while the LM is not as accurate at predicting the critical point.

Quantitative assessments of the predictive properties of these three models can be obtained by evaluating accuracy and error.
Accuracy is defined as the percentage of correctly predicted samples; higher accuracy means better prediction ability.
We measure the error of the fitting by the cross entropy~\cite{goodfellow2016deep}, which measures the closeness of two probability distributions $p$ and $q$. The cross entropy is defined as (for discrete distributions)
\begin{equation}
H(p,q) = -\sum_x p(x)\log q(x),
\end{equation}  which is the expectation value of $-\log q(x)$ for the random variable $x$ following distribution $p$. Here $p$ is the true probability distribution and $q$ is the predicted probability distribution. 
For our problem, the distribution is discrete and has only two cases, topological and trivial, so the cross entropy reduces to just the log loss function for a binary classification problem
\begin{equation}
L(y,\hat y) = \frac{1}{n}\sum_{i=1}^n [- y_i\log \hat y_i - (1-y_i) \log (1-\hat y_i) ],
\end{equation}
where $y_i$ is the true probability of being in the topological phase, and $\hat y_i$ is the predicted probability.
Small errors indicate a better model, and indeed the LM and NN models are trained on training data to reduce the error. However, the the RF model is trained based more on accuracy. 

The accuracies of the LM, NN, and RF models on the test data are 0.923, 0.971, and 0.978, respectively.
The corresponding errors are 0.202, 0.277, and 0.106.
The LM has the lowest accuracy among the three, due to its simple linear assumption, while NN and RF models perform similarly.
Nevertheless, we emphasize the use of the RF method has some advantages including the fact that it captures a high level of accuracy
while requiring much fewer parameters than the NN model.

Another benefit of the RF model is the ability to interpret how it makes classification decisions. To illustrate this we can plot the feature importance of the model (Fig.~\ref{fig:importance}). 
Feature importance measures the number of splits in a tree that includes the feature~\cite{breiman1984classification},
and higher feature importance means that the feature is more influential on prediction results.
We use the ES as features in our model.
	For each band in the ES the feature importance is calculated (these bands are the ES states near $0$, $1$, and the two in the mid-gap region. 
	To better understand the roles played by each band, we put Fig.~\ref{fig:importance} and Fig.~\ref{fig:ces} side-by-side.
	The vertical axis of Fig.~\ref{fig:importance} is the same as Fig.~\ref{fig:ces} with the bands represented by their averaged position from Fig.~\ref{fig:ces}.
	The horizontal axis shows the importance of each feature.
	As shown in the figure, the middle two bands of the ES have the highest influence on predictions, which
	strongly indicates that the RF model focuses on the degeneracy of the ES to perform its classification decisions.
Similar feature properties were also observed for our trained LM model, i.e.,  the coefficients of most features are nearly zero except for the features in the middle of the ES. We note that it is difficult to interpret coefficients of the NN model, so we have little that we can interpret about its behavior.

\begin{figure}
	\centering
	\includegraphics[width=0.9\linewidth]{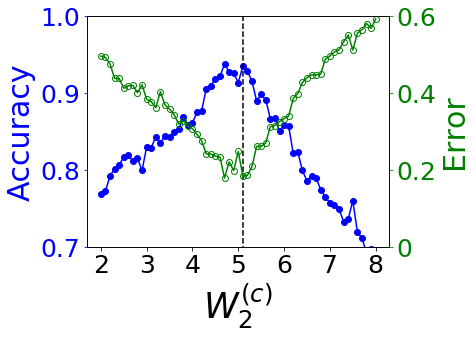}
	\caption{ Accuracy and log loss error of predictions for the RF model with different test choices for transition points $W_2^{(c)}$.
	The point with the highest accuracy (lowest log loss) agrees well with the true transition point as anticipated from the confusion scheme.}\label{fig:ces_fit_confusion}
\end{figure}

Let us move on to see how these methods can be extended. We trained our models with data deep in the topological and trivial phases, but a prior knowledge of the approximate phase transition point was needed to determine a reasonable parameter region to produce the training data. It would be more ideal if the RF model could also be used to find an unknown critical point as well. Indeed, this is possible if we use a scheme similar to the confusion scheme~\cite{van2017learning}. 
This method is a trial-and-test scheme that finds a point where two regions can be best distinguished by the model. 
This point is then the phase transition point.
To use this scheme we choose a possible transition point at $W_2^{(c)}$ and calculate the prediction accuracy and error.
If $W_2^{(c)}$ is the true transition point, we will produce high accuracy and low error. Otherwise the accuracy will be low and error will be high.
We plot the accuracy and error at different values of $W_2^{(c)}$ in Fig.~\ref{fig:ces_fit_confusion} for the RF model.
The $\wedge$ shape of the accuracy and the $\vee$ shape of the error are consistent with each other.
These results suggest that the transition point is $W_2\approx5$, which is consistent with the analytic result indicated by the vertical dashed line. This shows that in principle we could have employed the confusion scheme to find the disorder driven critical point in order to begin our model training, instead of knowing the exact critical point from an analytic calculation.

\begin{figure}
	\centering
	\subfigure[]{\includegraphics[width=0.49\linewidth]{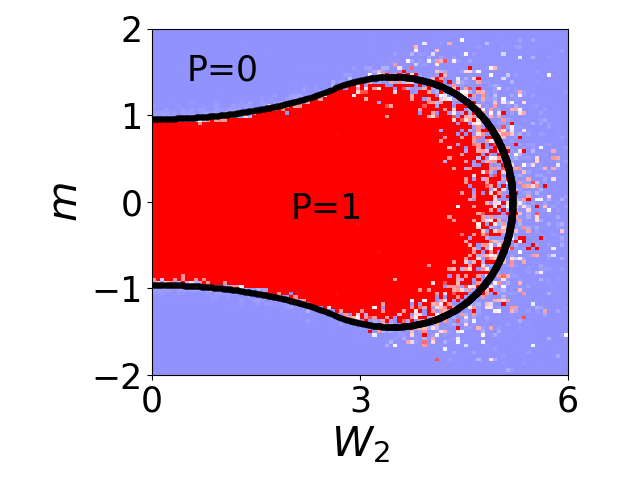}\label{fig:phase1}}
	\subfigure[]{\includegraphics[width=0.49\linewidth]{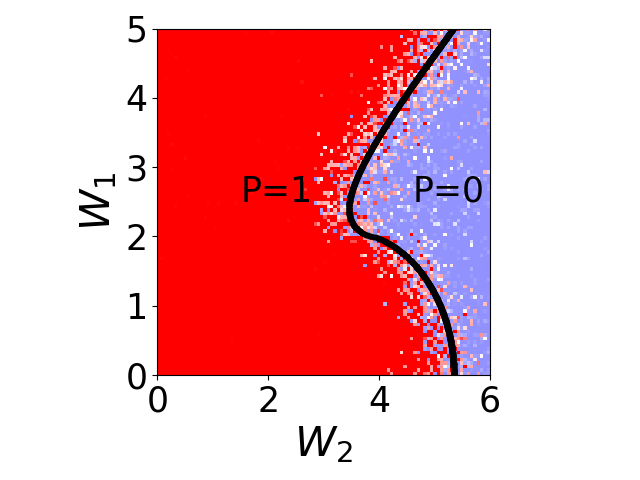}\label{fig:phase2}}	
	\caption{The predicted phase diagram of (a) $W_1=1.0$ and (b) $m=0.5$. The black solid lines are theoretical phase boundaries.
	$P$ is the predicted probability of being in the topological phase.}\label{fig:phase}
\end{figure}

So far we have found that the trained model can locate the transition point with relatively high accuracy, 
even though the model is not given training data information near the phase boundary. Now we want to see if we can expand the region of applicability of our model to a wide range of phase space. For example, we can take two other cross sections of the three-dimensional phase diagram: (i) fixed $W_1=1$ with varying $(W_2, m)$ or (ii) fixed $m=0.5$ with varying $(W_2, W_1).$
The phase diagrams for these cross-sections are plotted in Fig~\ref{fig:phase}a,b respectively.
The colormap indicates the predicted probability of being in the topological phase, and the exact phase boundaries are plotted as solid black lines. 
As can be seen, the RF model makes predictions with high confidence deep in the phases. 
When disorder is small, the predicted phase boundaries match very well with the exact ones,
while there are some deviations near the phase boundaries at large disorder. This gives us confidence that our model generalizes to a broader range of parameters than those on which it has been trained. 

\begin{figure}
	\centering
	\includegraphics[width=0.9\linewidth]{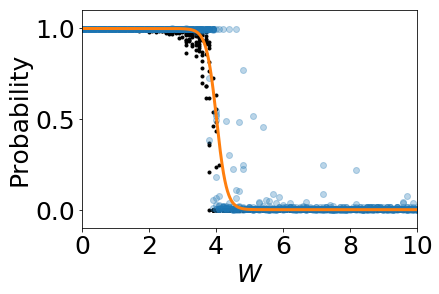}
	\caption{The predicted probability of being in the topological phase of Kitaev model in symmetry class BDI.}\label{fig:ces_fit_k}
\end{figure}

Since the robust properties of the ES are characteristic features in topological phases, especially in 1D, we expect that the 
RF algorithm that we trained  for the model in Eq.~\eqref{eq:chiral}
can be applied to a much broader set of symmetry-protected topological phases. Furthermore, since the stable features of  
the ES - used to train our machine learning algorithm - rely on global symmetries of the system, we expect to be able to
observe and quantify the breakdown of the method once symmetry-breaking effects are present.

As an example of the former, let us test the applicability of the RF model to another system. 
As an example, we apply our class AIII trained RF model to a disordered fermionic Kitaev chain in class BDI, whose  
Hamiltonian is given by~\cite{kitaev2001unpaired}
\begin{equation}
H = \sum_n [ t_n \;ib_n a_{n+1} + m_n \;ia_n b_n ],
\end{equation}
where $a_n$ and $b_n$ are Majorana fermions.
We add disorder to the parameters $t_n=1+W_1\omega_1$ and $m_n=m+W_2\omega_2$ 
where $t_n$ and $m_n$ here can be interpreted as inter-cell and intra-cell Majorana coupilng terms. 
In the clean limit corresponding to $W_1 = W_2 =0$, this model is a one-dimensional topological superconductor
with one (zero) isolated Majorana end state at each end for $|t| >|m|$ ($|t| <|m|$). 
For the test region we chose the line $W=2W_1=W_2$ and $m=0.5$.
The prediction results for this model are shown in Fig.~\ref{fig:ces_fit_k}, where, similar to Fig.~\ref{fig:ces_fit}, the dots represent predicted probabilities of being in the topological phase.
The orange curve is fitted using Eq.~\ref{eq:fit}, and we find a predicted phase transition near $W=4$.
The black dots are winding numbers calculated in real space for the same system~\cite{ian2014topological}. 
The transition point determined by the two methods are consistent with each other.

\begin{figure}
	\centering
	\subfigure[]{\includegraphics[width=0.49\linewidth]{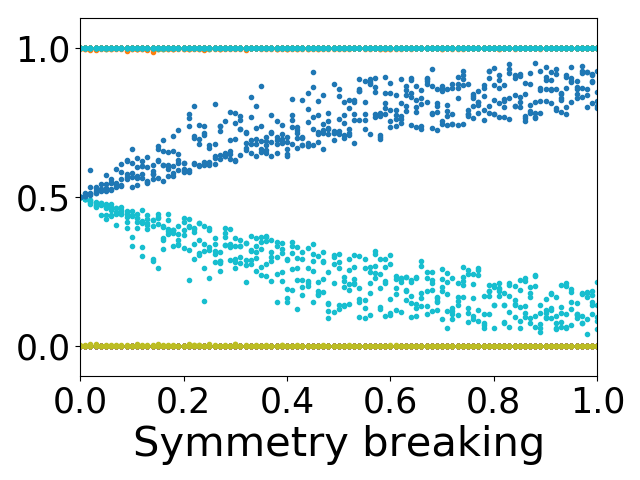}\label{fig:pred_break_es}}
	\subfigure[]{\includegraphics[width=0.49\linewidth]{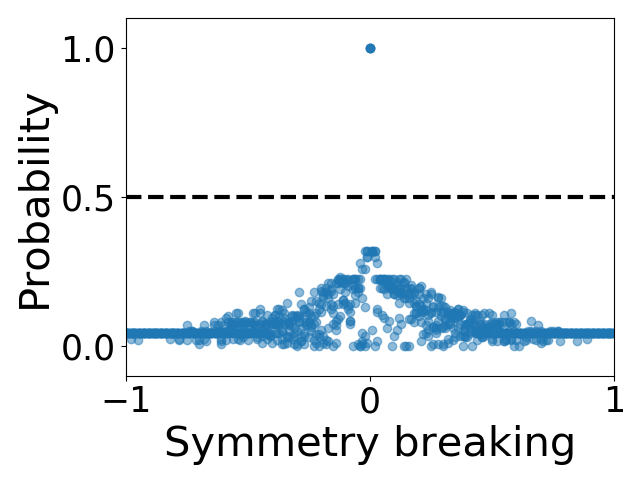}\label{fig:pred_break}}
	\caption{(a) Entanglement spectrum with chiral symmetry breaking term. 
		(b) The probability of being in the topological phase with symmetry breaking term added to the system.
		The model predicts that all configurations are in the trivial phase except when symmetry is preserved.}
\end{figure}

Now let us try to characterize some effects of symmetry breaking for the AIII model. To illustrate this  we add a small $\sigma_z$ term that breaks the chiral symmetry that protects the phase. 
We fix the other parameters as $m=0.5$ and $W_1=W_2=1$, so that the system is in topological phase when the symmetry is not broken, and 
add a term proportional to  $\sum_n c^{\dagger}_n \sigma_z c_n$ to Eq. \ref{eq:chiral}. Since the topological phase is protected by chiral symmetry this term immediately breaks down the topological phase and the degeneracy in the entanglement spectrum is lifted.
As the symmetry breaking term becomes stronger, the hitherto degenerate states in the middle of the entanglement spectrum 
are split away from each other, as indicated in Fig.~\ref{fig:pred_break_es}.
The resulting predictions made by the RF model are shown in Fig.~\ref{fig:pred_break}, where the blue dots are the raw prediction probabilities. 
In the absence of symmetry breaking terms, the model confidently predicts a topological phase for these model parameters. 
On the other hand, as soon as the symmetry breaking strength is non-zero, the predicted probability immediately drops below 0.5,
showing the sensitivity of the RF model to the removal of the degeneracy in the ES. Furthermore,
this probability goes down gradually as the symmetry breaking strength increase.

\begin{figure}
	\centering
	\subfigure[]{\includegraphics[width=0.49\linewidth]{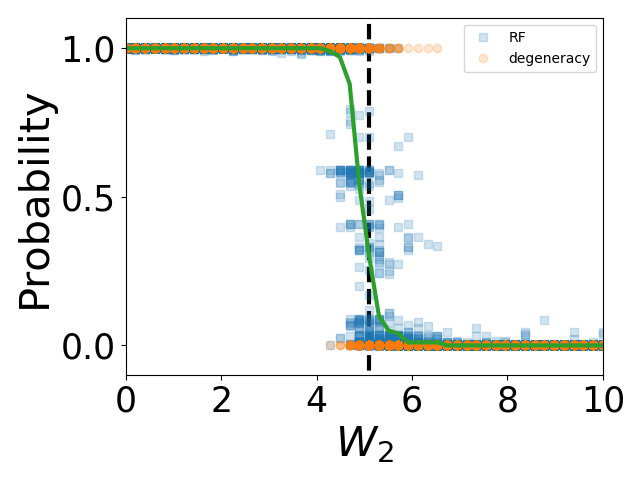}\label{fig:cee_disorder}}
	\subfigure[]{\includegraphics[width=0.49\linewidth]{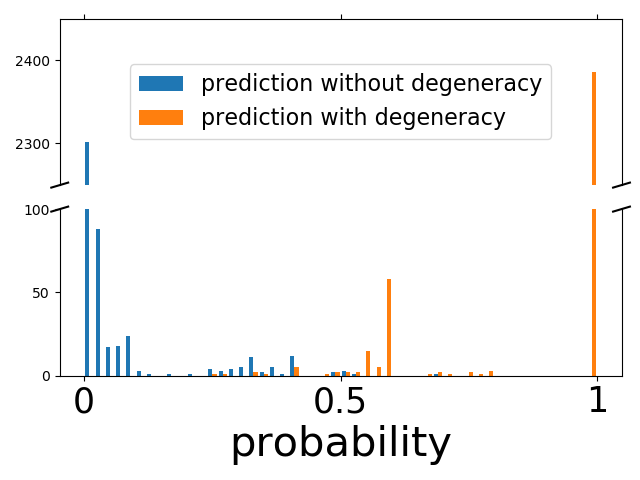}\label{fig:ana_pred}}
	\caption{(a) Predictions using the simple classifier that checks for degeneracy in the entanglement spectrum (orange dots) compared with random forest classifier (blue squares). 
		The green line is average of predictions for the degeneracy method. The analytically calculated phase transition point is indicated by a vertical dashed line.
	(b) Distribution of predictions of the random forest model for two different testing data sets. 
	When the test entanglement spectrum data has (no) double degeneracy, most predictions give probability one (zero) as being in the topological phase.}
\end{figure}

Finally, we can compare the predictions of the trained RF model with the very simple classifier of just checking for the degeneracy of the mid-gap entanglement modes. To use the entanglement degeneracy to make predictions of the phase we classify our data using a threshold and associate gaps in the ES smaller than 0.001 to the topological phase with probability 1, and larger gaps to be in the trivial phase with probability 1.
We find that the predicted results using this simple method give accuracy 0.977, which is close to the accuracy of our random forest classifier.
We plot the predictions from degeneracy in Fig.~\ref{fig:cee_disorder} with orange dots. 
Since we can only predict one or zero, we take the average of the predictions as the probability (green line).
The predictions of the random forest model are shown with blue squares for comparison and the two predictions match extremely well.

We can dig a bit deeper into understanding how the RF model is classifying based on the ES degeneracy. We plot the distribution of random forest predictions in Fig.~\ref{fig:ana_pred} for two sets of testing data: one has degeneracies in the ES while the other does not.
Note that the y-axis is cut in the middle to reveal details for smaller $y$ values.
From the figure we can see that if the ES has degeneracy the RF model almost always predicts the state as topological (probability$>0.5$),
while for non-degenerate ES, the model predicts the states as trivial (probability$<0.5$) in most cases.
Among the 3\% incorrect predictions made by either the RF model or the simple degeneracy classifier, 
84\% are wrong by both models; 
9\% are wrong by the RF model but correct by degeneracy;
and 7\% are correct by the RF model but wrong by degeneracy.
Therefore, the RF predictions are consistent with the predictions by degeneracies. 
So, after careful investigation we find that the RF model is making predictions based on the mid-gap degeneracy of the ES. 
We expect that the training process is essentially finding the best threshold for the gap size.
The threshold ends up being about 0.0015, which is close to the value we set for our simple degeneracy classifier.

\section{Summary}
In summary, we applied the random forest  model to classify disordered topological phases. 
Compared with the linear model, random forest gives better predictions. 
On the other hand, it preserves the easy interpretability of the linear model as compared to neural networks.
Because of the generality of the entanglement spectrum, the model trained on a small training dataset can be generalized to test data
in a larger phase space, and even to other models in different symmetry classes. 
A closer look at the RF model indicates that the model is capturing the degeneracy of the mid-gap entanglement spectrum modes and is very sensitive to any symmetry breaking which splits the modes. 

{\bf{Acknowledgements:}}
This article is presented as a memorial tribute for Shou-Cheng Zhang and combines two of his recent interests of topological insulators and machine learning. TLH thanks him for the advice and support given during their long period of collaboration. YZ and TLH acknowledge support from the US National Science Foundation under grant DMR 1351895-CAR. LHS is supported by a faculty startup at Emory University.
 
\bibliographystyle{unsrt}
\bibliography{refs}{}

\end{document}